\documentstyle[12pt,fleqn]{article}
\date{}
\newcommand{\nl}{\nonumber\\}
\newcommand{\be}{\begin{equation}}
\newcommand{\ee}{\end{equation}}
\newcommand{\ba}{\begin{eqnarray}}
\newcommand{\ea}{\end{eqnarray}}

\newcommand{\la}[1]{\label{#1}}

\begin{document}
\begin{flushright}
hep-ph/9705364
\end{flushright}
\renewcommand{\thefootnote}{\fnsymbol{footnote}}
\vspace{1.5cm}

\noindent{\Large\bf Determination of Scalar Meson Masses\\
in QCD-inspired Quark Models\footnotemark[1]}

\vspace{0.5cm}

\noindent A. A. Andrianov\footnotemark[2]\footnotemark[3] 
and V. A. Andrianov\footnotemark[2]\\

\medskip

\noindent
{\small Laboratory for Particle and Nuclear Theory,
Sankt-Petersburg State University,\\
198904 Sankt-Petersburg, Russia\\
E-mail: andrian@snoopy.phys.spb.ru}

\vspace{1.cm}

\noindent{\bf Abstract.} We compare two QCD-inspired quark models
with four-fermion interaction, without and with the remnant coupling
to low-energy gluons, in the regime of dynamical chiral symmetry
breaking (DCSB). The first one, the Nambu-Jona-Lasinio (NJL)  model
ensures the factorization of scalar and pseudoscalar meson poles in
correlators, the well-known Nambu relation between the scalar meson
mass and the dynamical quark mass, $m_{\sigma} = 2 m_{dyn}$,
and the residual chiral symmetry in coupling constants
characteristic for the linear $\sigma$-model. The second one, the
Gauged NJL model, happens to be qualitatively different from the NJL
model, namely, the Nambu relation is not valid and the factorization of
light meson poles does not entail the residual chiral symmetry,
i.e. it does not result in a linear $\sigma$-model. The more complicated
DCSB pattern in the GNJL model is fully explained in terms of excited
meson states with the same quantum numbers. The asymptotic restrictions
on parameters of  scalar and pseudoscalar meson states are derived from
the requirement of chiral symmetry restoration at high energies.

\setcounter{footnote}{1}
\footnotetext{This is a translation of the paper published in:
Zapiski Nauch. Sem. PDMI (Proc. Steklov Math. 
Inst., St.Petersburg Dept., Russia),
v.245/14 (1996) p.5 (Eds. L.D.Faddeev, P.P. Kulish and A.G.Izergin, 
to appear as an AMS translation)}
\setcounter{footnote}{2}
\footnotetext{
Supported by  RFBR (Grant No. 95-02-05346a),
and by  INTAS (Grant No. 93-283ext).}
\setcounter{footnote}{2}
\footnotetext{
Supported by GRACENAS
(Grant No. 95-06.3-13).}

\newpage
\noindent {\bf 1. Introduction}

\bigskip

\noindent The QCD-inspired quark
models with attractive four-fermion interaction are often applied for the
truncation of the low-energy QCD in the hadronization regime [1-9].
In the common approach [1,2] the local
four-fermion interaction only is involved to induce
the DCSB due to strong attraction in
the scalar channel. As a consequence, the dynamical quark mass
$m_{dyn}$ is created, massless pions (in the chiral limit, $m_{current} = 0$)
and the massive scalar meson with the
mass, $m_{\sigma} = 2 m_{dyn}$ arise and
the residual chiral symmetry characteristic for the linear $\sigma$-model [2]
holds in coupling constants (see below).
On the other hand the ef\/fective QCD
action for quarks at low energies
contains, in general, the remnant interaction to low-energy, background
gluons  which is responsible for the confinement and
hadron
properties at intermediate energies, in particular, for the formation
of excited meson states [10]. Therefore, when gluons
are treated  as being low-energy, background ones
the Gauged NJL model may be thought of
as a better truncation of the low-energy QCD effective action [5].
As the DCSB holds also for
such a model the appearance of large dynamical quark mass gives rise the
possibility to calculate nonperturbative gluon corrections for
hadron parameters in the condensate approach [3-6].

When comparing two
models in the description of hadron properties one has to examine
to what extent
the Gauged NJL model leads to the same hadron lagrangian
as the conventional NJL one, after proper renormalization of
its basic characteristics [4]. It is one of the aims of our paper to show that
two models are qualitatively different since the coupling to background
gluons breaks the linear $\sigma$-model relations  for masses and coupling
constants which are valid in the non-gauged case.

In order to explain the roots of this difference we consider
the qualitative features of QCD hadron spectrum
in the planar limit (large $N_c$). In this approach the correlators
of color-singlet quark currents are saturated by narrow meson resonances.
In particular, the two-point correlators of scalar and pseudoscalar
quark densities are represented by the
sum of related meson poles,
\ba
\Pi^S (Q) &=&
- \int d^4x \,\exp(iqx)\ <T\left(\bar\psi\psi(x) \,\, \bar\psi
\psi(0)\right)>_{planar}\nl
&=&\,
\sum_n \,\frac{Z^S_n}{Q^2 + M^2_{S,n}} + C^S_0 + C^S_1 Q^2;\nl
\Pi^P (Q) \,&=&\,
 \int d^4x \,\exp(iqx)\ <T\left(\bar\psi\gamma_5\psi(x) \,\,
 \bar\psi\gamma_5\psi(0)\right)>_{planar}\nl
\,&=&\,
\sum_n \,\frac{Z^P_n}{Q^2 + M^2_{P,n}}  + C^P_0 + C^P_1 Q^2,
\la{planar}
\ea
in the Euclidean momentum region, $q^2 = - Q^2 < 0$.
The absence of physical thresholds for $q^2 > 0$
in the large $N_c$ limit is ensured
by the confinement of color intermediate states.
The high-energy asymptotics is controlled [11] by the
perturbation theory and the operator product expansion due
to asymptotic freedom of QCD. Thereby the same
correlators have the power-like behavior at large $Q^2$,
\be
\Pi^S(Q) |_{Q^2 \rightarrow \infty} \,\simeq
\Pi^P(Q) |_{Q^2 \rightarrow \infty}
\,\simeq \,\frac{N_c}{8\pi^2} Q^2\,\ln\frac{Q^2}{\mu^2}.
\ee
When comparing (1) and (2) one concludes that the infinite series of
resonances with the same quantum numbers should exist in order to reproduce
the perturbative asymptotics. Besides, one can prove [12] that
\be
\left(\Pi^S(Q) -  \Pi^P(Q)\right)_{Q^2 \rightarrow \infty} =
O \left(\frac{1}{Q^6}\right) ,\la{CSR}
\ee
and the chiral symmetry is  restored at high energies.

Thus the QCD-inspired quark models are expected to reproduce the part of
QCD meson spectrum in the planar limit
and the asymptotic chiral symmetry restoration for higher energies.
In particular, the conventional NJL model describes one scalar and one
pseudoscalar (multiplet of) state whereas the Gauged NJL model should
contain additional excited meson states with the same quantum numbers.
Just the appearance of additional poles in correlators makes two
models to be different in the DCSB pattern.

The rest of the paper contains the derivation of two-point
correlators in the NJL and Gauged NJL models, the proof of the
violation of linear $\sigma$-model relations and the soft-momentum
determination of scalar meson parameters. At the end the
two-resonance ansatz is performed to saturate the deviation
of NJL and GNJL models. The asymptotic sum rules based on
the chiral symmetry restoration  are obtained for the resonance
characteristics.

\bigskip

\noindent{\bf 2. The definitions}

\bigskip

\noindent We start from the NJL model without gluons and
introduce the auxiliary
scalar, $\widetilde\sigma = \sigma(x) + m_{dyn};\quad (m_{dyn}\equiv m)$ and
pseudoscalar, $\pi(x)$ fields.
Respectively the basic lagrangian density
can be presented in two forms (the euclidean-space formulation is taken here),
\ba
{\cal L}(x))\,&=& \,\bar\psi \left(i \widehat\partial
 + iS(x) + \gamma_5 P(x)\right)\psi\,+\,\frac{g^{2}}{4N_{c}}\,
\left[(\bar\psi \psi)^{2} - (\bar\psi\gamma_5 \psi)^{2}\right]\nl
&\Longleftrightarrow &\,\bar\psi\bigl(\widehat D\, + iS(x) +
\gamma_5 P(x) \bigr)\psi\,
+\frac{N_{c}}{g^2}\,\left((\sigma (x) + m)^2 + \pi^2(x)\right),
\ea
where $\psi \equiv \psi_{i}$
stands for color fermion f\/ields with $N_{c}$ components, $S(x), P(x)$ are
external scalar and pseudoscalar sources and
the Dirac operator is
\be
\widehat D = i \widehat\partial + im +
i\sigma(x) + \pi(x)\gamma_5;\quad \widehat\partial \equiv i\gamma^{\mu} .
\partial_{\mu} \la{D}
\ee
The vector and axial-vector fields are not included
since the main story is interplayed between scalar and
pseudoscalar channels. We simplify our analysis setting up the
number of f\/lavours $N_{F}=1$ and the current quark mass $m_{q}=0$.

Let us make the change of fields, $\sigma \rightarrow \sigma - S;\quad
\pi \rightarrow \pi - P$. Then
the regularized fermion vacuum functional $Z^{F}_{reg}$ is described by
the following relations,
\begin{equation}
\ln Z^{F}_{reg}(\sigma, \pi)\,=\,\ln \,\biggl\langle \exp\biggl(
-\int\! d^{4}x\,{\cal L}_{F}(\sigma(x), \pi(x))\biggr)\biggr\rangle_
{\bar\psi\psi} \,=\,  N_c \mbox{\rm Tr} \ln\widehat D\vert_{reg}
\end{equation}
The corresponding effective action for auxiliary fields reads
\ba
S_{eff} &=&- N_c \mbox{\rm Tr} \ln\widehat D\vert_{reg}\nl
&& + \frac{N_c}{g^2} \int\! d^{4}x\,
(\sigma^2 + 2m\sigma  + \pi^2 -2\sigma S- 2m S -
2\pi P + S^2 + P^2 + m^2 ) \la{seff}
\ea
where a momentum-cutoff
regularization is implied.
The parameter $g$ is a four-fermion constant. When relating to
the BRZ denotations [8] it happens to be $g^2 = 8\pi^2 G_s/\Lambda^2$.

\bigskip

\noindent{\bf 3.  Parameters of scalar and pseudoscalar mesons in
the conventional NJL model}

\bigskip

\noindent As usual the dynamic mass is subject to the mass-gap equation
in which $\Lambda$ is a  $O(4)$-invariant cutoff,
\be
m (\Lambda^2 - \frac{8\pi^2}{g^2} - m^2\ln\frac{\Lambda^2}{m^2}) +
O(\frac{m^2}{\Lambda^2}) = 0 .\la{l1}
\ee
Herein and further on we omit the terms of order $1/\Lambda^2$ which
do not change the result drastically. This equation provides
the cancellation of the linear in
$\sigma$ terms in (\ref{seff}),
with the tadpole in the fermion determinant,
\ba
\frac{2N_c m}{g^2} &=& i <\bar\psi \psi> =\frac{i  N_c}{(Vol.)}
\mbox{\rm Tr}\left(\widehat D^{-1}\right)_{reg} |_{\sigma,\pi = 0} = \nl
&=& 4 N_c \int_{p < \Lambda} \frac{p^2 d(p^2)}{16\pi^2}\frac{m}{p^2 + m^2} .
\la{tad}
\ea
The quadratic in fields part of $S_{eff}$ determines the inverse propagators
("kinetic terms") for meson fields,
\be
S_ {eff} = \frac12\int \frac{d^4Q}{(2\pi)^4}
\left[ \sigma(Q)\Gamma_{\sigma}(Q^2) \sigma(-Q) +
\pi(Q)\Gamma_{\pi}(Q^2)\pi(-Q)\right], \la{S1}
\ee
where  in the simple NJL model without gluons one has,
\ba
\Gamma_{\sigma} = \frac{2N_c}{g^2} -N_c \int_{p<\Lambda}
\frac{d^4p}{(2\pi)^4}\mbox{\rm tr}\Delta (\frac{Q}{2})\Delta
(\frac{-Q}{2}),\nl
\Gamma_{\pi} = \frac{2N_c}{g^2} -N_c\int_{p<\Lambda} \frac{d^4p}{(2\pi)^4}
\mbox{\rm tr}i \gamma_5 \Delta (\frac{Q}{2}) i\gamma_5\Delta (\frac{-Q}{2}),
  \la{l2}
\ea
and the fermion propagator is given by
$\Delta(Q/2) = ( \widehat p + \widehat Q/2 + im)^{-1}$. The correspondence
to the BRZ denotations [8] is following,
\ba
\bar\Pi_S = \frac{1}{g_s} - \Gamma_{\sigma};\quad
\quad\bar\Pi_P = \frac{1}{g_s} - \Gamma_{\pi};\nl
\frac{1}{g_s} =\frac{N_c \Lambda^2}{4\pi^2G_s}
= \frac{2N_c}{g^2} = -
\frac{<\bar q q>}{m} = i \frac{<\bar\psi \psi>_{E}}{m}.
 \la{BRZ}
\ea
Both integrals in (\ref{l2}) are divergent. If we take into account the
mass gap equation then they are logarithmically divergent only.
Let us evaluate the integrand in (\ref{l2}),
$$\mbox{\rm tr}(1, i\gamma_5)\left(\widehat p +\frac{\widehat Q}{2} +
im\right)^{-1}(1, i\gamma_5) \left(\widehat p -\frac{\widehat Q}{2} +
im\right)^{-1}$$
$$= 4\left( (p_{\mu} +\frac{Q_{\mu}}{2})^2 + m^2\right)^{-1}\left(p^2
-\frac{Q^2}{4}\mp m^2 \right)
\left( (p_{\mu}-\frac{Q_{\mu}}{2})^2  +m^2\right)^{-1}$$
$$= 2\left[\left( (p_{\mu} +\frac{Q_{\mu}}{2})^2 + m^2\right)^{-1} +
\left( (p_{\mu} -\frac{Q_{\mu}}{2})^2  + m^2\right)^{-1}\right]$$
\be
- 2 \left( Q^2 + 2m^2 \pm 2m^2 \right)
\left( (p_{\mu} +\frac{Q_{\mu}}{2})^2 + m^2\right)^{-1}
\left( (p_{\mu}-\frac{Q_{\mu}}{2})^2  +m^2\right)^{-1}\la{l99}
\ee
The sum of demoninators is, $2 p^2  +2 m^2 + (Q^2/2)$ and after its
separation in the numerator of Eq.(\ref{l99}) one arrives to the required
decomposition of loop integrals. Two first terms at zero-momentum represent
the tadpole integrals in (\ref{tad}) and therefore precisely cancel the
first terms in (\ref{l2}). This is a work of the mass-gap eq.(\ref{l1}). Thus,
depending on the regularization we derive
the inverse $\sigma$-model propagators,
\ba
\Gamma_{\sigma}&=&(4m^2 + Q^2) I(Q^2) + \xi Q^2 + O(\frac{1}{\Lambda^2})
\equiv (m_{\sigma}^2 + Q^2) I_{\sigma}(Q^2),\nl
\Gamma_{\pi} &=& Q^2 I(Q^2)  + \xi Q^2 + O(\frac{1}{\Lambda^2})\equiv  Q^2
 I_{\pi}(Q^2),\la{l3}
\ea
where the parameter $\xi$ is a finite constant  characteristic to
the regularization, for instance, in the symmetric cutoff
regularization (\ref{l2}) one has $\xi = N_c/32 \pi^2$. Respectively
the decay formfactor is provided by
the last term in (\ref{l99}),
\be
I (Q^2)= 2N_c \int_{p < \Lambda} \frac{d^4p}{(2\pi)^4}\frac{1}{(p +
\frac12 Q)^2 + m^2}\frac{1}{(p - \frac12 Q)^2 + m^2}.\la{l4}
\ee
It is regularization independent up to a redefinition
of $\Lambda$ when neglecting the orders of $1/\Lambda^2$.
For a convenience we display the analytic representation
for $I(Q^2)$,
\ba
I(Q^2)&=& \frac{N_c}{8\pi^2}\left( \ln\frac{\Lambda^2}{m^2} - 1
-\int_{0}^{1} dx \ln \left(1 + x(1 - x)
\frac{Q^2}{m^2}\right)\right)\nonumber\\
&=& \frac{N_c}{8\pi^2}\left( \ln\frac{\Lambda^2}{m^2} + 1
+ \sqrt{1 + \frac{4m^2}{Q^2}}
\ln\frac{\sqrt{1 + \frac{4m^2}{Q^2}} - 1}{\sqrt{1 + \frac{4m^2}{Q^2}} + 1}
\right)
\ea
for euclidean momenta $Q^2 > 0$.
If one defines the regularization where $\xi = 0$ (or an appropriate
subtraction of $\xi$-term in the arbitrary regularization) then one
arrives to the residual chiral symmetry and thereby to the generalized
linear $\sigma$-model,
\be
I_{\sigma}(Q^2) = I_{\pi}(Q^2) \equiv I (Q^2).\la{l5}
\ee
Then as a consequence of the NJL-mechanism of DSB we have
precisely $m_{\sigma} = 2 m$.
The generating functional for scalar and pseudoscalar
correlators can be derived after one integrates over $\sigma, \pi$
variables keeping
in the effective action the quadratic terms
only (the main large-$N_c$ contribution). The gaussian integral is calculated at
the extremum,
\ba
\sigma_c (Q) &=&  \frac{2N_c}{g^2} \Gamma_{\sigma}^{-1}
S(Q) =\frac{2N_c}{g^2 I(Q)} \frac{1}{Q^2 + 4 m^2} S(Q)\nonumber\\
\pi_c (Q) &=&  \frac{2N_c}{g^2} \Gamma_{\pi}^{-1} P(Q)
=\frac{2N_c}{g^2 I(Q)} \frac{1}{Q^2} P(Q)
\ea
and results in the generating functional,
$ W = - \ln Z(S, P)$,
$$ W^{(2)} = \int \frac{d^4Q}{(2\pi)^4}
\left[ \frac12 S(-Q) \left(\frac{<\bar\psi\psi>^2}{m^2}  \Gamma_{\sigma}^{-1}
 - i\frac{<\bar\psi\psi>}{m} \right)S(Q)\right.$$
\be
\left. +  \frac12 P(-Q) \left(\frac{<\bar\psi\psi>^2}{m^2}
 \Gamma_{\pi}^{-1} - i\frac{<\bar\psi\psi>}{m} \right)P(Q)-
S(0) (2\pi^4) \delta^{(4)}(Q) i <\bar\psi\psi>\right] .\label{Wact}
\ee
The calculation of two-point correlators correspond to the second derivative
of this action with subtraction of disconnected product
of condensates (it corresponds to elimination of the last term $\sim S(0)$
in above Eq.).
$$\int d^4x \exp(iQx) \left[<T (\bar\psi\psi(x) \bar\psi\psi(0))>
- <\bar\psi\psi><\bar\psi\psi>\right]  $$
$$=- \int d^4x \exp(iQx)\frac{\delta^2 }{\delta S(x) \delta S(0)}\left(W^{(2)}
+ \int d^4y S(y) i <\bar\psi\psi>\right). $$
It is saturated by  first terms in (\ref{Wact}) quadratic in $S(x)$.
One has the scalar-meson pole  and another constant term which
plays some role [8] when one reduces
linear sigma model to the chiral one.
Similarly we obtain the pseudoscalar correlator which contains
also the pole and the constant term. One can compare them to
the correlators (124)-(127) and (152) of the BRZ
paper and
find the one-to-one correspondence if to go back to the Minkowski space
where
$-i <\bar\psi\psi>_E \rightarrow <\bar\psi\psi>_M \equiv <\bar q q>$.
The chiral symmetry restoration at high energies is ensured by (\ref{l5}).

\bigskip

\noindent{\bf 4. Soft-momentum predictions}

\bigskip

\noindent
Let us now develop the soft-momentum expansion of above
kinetic terms,
\be
\Gamma^0_{\sigma}
 =a_0^{\sigma} + Q^2 a_1^{\sigma} + \cdots,\quad
\Gamma^0_{\pi} = Q^2 a_1^{\pi} + Q^4 a_2^{\pi} + \cdots.\la{sm1}
\ee
The coefficients $a_j^{\sigma,\pi}$ are related to the Taylor expansion
of (\ref{l3}),
\ba
a_0^{\sigma} &=& m^2_{\sigma} I_{\sigma}(0),\quad a_1^{\sigma}
= I_{\sigma}(0) + m^2_{\sigma} I'_{\sigma}(0) \nl
a_1^{\pi} &=& I_{\pi} (0),\quad
a_2^{\pi} = I'_{\pi} (0)\la{l6}
\ea
We see that in the vicinity of $Q^2 \sim 0$ the would-be
scalar-meson decay constant $a_1^{\sigma}$ and  pion constant
$a_1^{\pi}$ do not coincide. Moreover their
difference represents the finite constant nearly independent on a
regularization, $a_1^{\sigma} - a_1^{\pi} = 4m^2 I'(0)$.
We remark that the corresponding integral $I'(0)$ is convergent and
in the limit $\Lambda \rightarrow \infty$
 takes the value
$I'(0) = -  N_c / 48 \pi^2 m^2$.
On the other hand, when the residual chiral symmetry (\ref{l5}) is present
(for $\xi = 0$) the latter one reveals itself in a series of $\sigma$-model
identities between the coefficients of soft-momentum expansion,
\be
\frac{a_0^{\sigma}}{a_1^{\pi}} =
m^2_{\sigma},\quad
a_1^{\pi} a_1^{\sigma}  = (a_1^{\pi})^2 + a_2^{\pi}
a_0^{\sigma},\ldots\la{l7}
\ee
We emphasize that the first relation is meaningful iff the next ones
are valid. But only if $I_{\sigma} = I_{\pi}$ one can use the first
relation in eq.(\ref{l7}) for the definition of the scalar meson mass.
Otherwise it is misleading. Obviously, in  the conventional NJL model
the scalar meson mass is easily determined from the
soft-momentum (derivative) expansion.  In the next section we shall see
that it is not the case for the Gauged NJL model.

\bigskip

\noindent{\bf 5. Gauged NJL model and status of $\sigma$-model}

\bigskip

The gauged NJL model is prepared conventionally
with help of the long derivative
in the Dirac operator, $i\partial_{\mu}\rightarrow D_{\mu}
=i\partial_{\mu} + G_{\mu}$ and $G_{\mu} \equiv  g_s T^a G^a_{\mu}$ is
a soft gluon field. The latter leads to the following modification of
quark propagators in loop integrals, eq.(\ref{l3}),
\be
\Delta (\pm\frac{Q}{2}) = \left(\pm\frac{\widehat Q}{2}+ i\widehat\partial
+\widehat G +im\right)^{-1} \la{p1}.
\ee
Respectively we examine the
 main large-$N_c$ contribution into the effective action given by averaging
of functionals $\Gamma_{\sigma,\pi} (\Delta)$ over gluons,
\be
\Gamma_{\sigma,\pi} \equiv <\Gamma_{\sigma,\pi}(\Delta)>_G. \la{l8}
\ee
We assume that the main DSB mechanism is due to
the strong four-fermion interaction and
soft gluons bring the corrections accumulated into condensates. Thereby we
imply that the derivative or large-mass expansion is reasonably good in
calculating the gluon contributions.

Let us analyze the general structure
of $\Gamma_{\sigma,\pi}$ which defines the inverse propagators for $\sigma$
and $\pi$-fields according to (\ref{S1}). For this purpose
we first evaluate the integrand in (\ref{l2}),(\ref{l8}),
\ba
&&\mbox{\rm Tr}(1,i\gamma_5)\left(\frac{\widehat Q}{2} +
i\widehat\partial +\widehat G +im\right)^{-1}(1, i\gamma_5)
\left(-\frac{\widehat Q}{2} + i\widehat\partial +\widehat G
+im\right)^{-1} =\nl
=&& \mbox{\rm Tr}\left( (\frac{Q_{\mu}}{2} + i\partial_{\mu}
+ G_{\mu})^2+ \widehat F + m^2\right)^{-1}\nl
&&\times\left(( i\partial_{\mu} + G_{\mu})^2  + \widehat F
-\frac{Q^2}{4}\mp m^2 + \frac12 [\widehat q,  \widehat D]\right)\nl
&&\times\left( (-\frac{Q_{\mu}}{2} + i\partial_{\mu} + G_{\mu})^2 + \widehat F
+m^2\right)^{-1} \la{l9}
\ea
where $\widehat F \equiv (1/4) [\gamma_{\mu},\gamma_{\nu}]
\cdot [D_{\mu}, D_{\nu}]$. The sum of demoninators is,
\be
\Sigma \equiv 2 (D_{\mu})^2  + 2\widehat F +\frac{Q^2}{2} +2 m^2
\ee
and after its separation in the second multiplier of product (\ref{l9})
one arrives to the following decomposition of loop integrals
(compare with eq.(\ref{l3})):
\ba
\Gamma_{\sigma} &=& (Q^2 + 4m^2) I_G(Q^2) + Q^2 \Xi(Q^2)
+O(\frac{1}{\Lambda^2}),\nl
\Gamma_{\pi} &=& Q^2 \left( I_G(Q^2) + \Xi (Q^2)\right)
+ O(\frac{1}{\Lambda^2}),
\la{l10}
\ea
where now the employed functions are defined as follows,
\ba
I_G (Q^2)&=& \frac12 \langle\mbox{\rm tr}<x|\left(\frac{Q_{\mu}}{2} +
i\partial_{\mu} + G_{\mu})^2+ \widehat F + m^2\right)^{-1}\nl
&&\times\left(
 -\frac{Q_{\mu}}{2} + i\partial_{\mu} + G_{\mu})^2+ \widehat F
+ m^2\right)^{-1} |x>\,\rangle_G;\nl
&=&\frac12 \int_{p < \Lambda} \frac{d^4p}{(2\pi)^4}\langle\mbox{\rm tr}
\left( (p_{\mu} +\frac{Q_{\mu}}{2} + i\partial_{\mu} + G_{\mu})^2
+ \widehat F + m^2\right)^{-1}\nl
&&\times\left( (p_{\mu} -\frac{Q_{\mu}}{2} + i\partial_{\mu} + G_{\mu})^2
+ \widehat F + m^2\right)^{-1}\rangle_G ;\nl
\Xi (Q^2) &=&\xi -\frac12 \langle \mbox{\rm tr}<x|\left(
(\frac{Q_{\mu}}{2} + i\partial_{\mu} + G_{\mu})^2+ \widehat F
+ m^2\right)^{-1}\nl
&&\times\frac{1}{Q^2} [\widehat q, \widehat D] \left((-\frac{Q_{\mu}}{2}
+i\partial_{\mu} + G_{\mu})^2 + \widehat F + m^2\right)^{-1}|x>\rangle_G \nl
&=& \xi -\frac12 \int \frac{d^4p}{(2\pi)^4}\langle\mbox{\rm tr}\left(
(p_{\mu} +\frac{Q_{\mu}}{2} + i\partial_{\mu} + G_{\mu})^2
+ \widehat F + m^2\right)^{-1}\nl
&&\times\frac{1}{Q^2} [\widehat q, \widehat p + \widehat D ]
\left( (p_{\mu} -\frac{Q_{\mu}}{2} +i\partial_{\mu} + G_{\mu})^2
+ \widehat F + m^2\right)^{-1}\rangle_G.\la{l11}
\ea
The integral for $\Xi$ is not divergent.
We have used the modified gap equation and the $1/\Lambda^2$ expansion
which results in the independence of parameter $\xi$ on gluon
condensates ( $\xi = N_c/32 \pi^2$ in the sharp cutoff regularization).
The representation for $I_{G}$ and $\Xi$ in Eqs.(\ref{l11}) contains
the trace over spin and color indices.
In the second, integral representation of above functions
\underline{the derivatives act on gluon fields} when one expands,
for example,  those functionals in the
perturbation series. In other words, we have to
calculate the mixed matrix element $<x|\cdots |k=0>.$
Evidently the relations of the linear $\sigma$-model (\ref{l5})
would survive if it were true that,
\ba
(Q^2  + 4m^2) I_G(Q^2) + Q^2 \Xi(
Q^2) &{?\atop=}& (Q^2 + m_{\sigma}^2) \left(I_G(Q^2) + \Xi(Q^2)\right) \nl
\mbox{or} \quad \Xi(Q^2) &{?\atop=}& \frac{4m^2 - m^2_{\sigma}}{m^2_{\sigma}}
I_G (Q^2).  \la{l12}
\ea
By comparing of integrals in (\ref{l11}) one can convince
oneself that for arbitrary momenta these functions
do not coincide, the equation (\ref{l12}) is
not valid and therefore the linear $\sigma$-model relations
(\ref{l5}), (\ref{l7}) are not fulfilled.
Thus we cannot effectively use the soft-momentum expansion for
the calculation of scalar meson mass and its decay constant.
Now let us consider the soft-momentum and large-mass expansion which leads
to the expansion in terms of gluon condensates. We stress that
the coefficients of large-mass expansion containing gluon condensates
are free of ultraviolet divergences and thereby they do not depend on
regularization if one neglects terms $O(1/\Lambda^2)$.
The new mass-gap equation takes the form,
\be
\Lambda^2 - \frac{8\pi^2}{g^2} - m^2\ln\frac{\Lambda^2}{m^2}
+ \frac{<G^2_{\mu\nu}>}{12N_c m^2} = 0.
\ee
First terms of above expansion for $\Gamma_{\sigma}, \Gamma_{\pi}$
read (see eq.(\ref{l6})) in the momentum-cutoff regularization after applying
the mass-gap equation:
\ba
a_0^{\sigma} &=& \frac{N_c}{2\pi^2}\left(m^2 \left(\ln
\frac{\Lambda^2}{m^2} - 1\right) + \frac{<G^2_{\mu\nu}>}{12N_c m^2}\right);\nl
a_1^{\sigma} &=& \frac{N_c}{8\pi^2}\left(\ln\frac{\Lambda^2}{m^2}
-\frac{5}{3} + \widetilde\xi - \frac{7<G^2_{\mu\nu}>}{120 N_c m^4}\right);\nl
a_1^{\pi} &=& \frac{N_c}{8\pi^2}\left(\ln\frac{\Lambda^2}{m^2} -1
+ \widetilde\xi + \frac{<G^2_{\mu\nu}>}{24 N_c m^4}\right);\nl
a_2^{\pi} &=& \frac{N_c}{48\pi^2 m^2}\left( - 1
- \frac{<G^2_{\mu\nu}>}{20 N_c m^4}\right) . \la{sme}
\ea
where $<G^2_{\mu\nu}> \equiv g^2_{QCD} <(G^a_{\mu\nu})^2>,\quad
\widetilde\xi = 1/4 = \xi\cdot 8\pi^2/N_c$.
One can verify explicitly that the linear $\sigma$-model relations
(\ref{l6}),(\ref{sm1}) are not fulfilled even after the
renormalization to the chirally symmetric point,
$a_1^{\sigma,\pi} \rightarrow a_1^{\sigma,\pi} - \xi; \quad
 \xi = N_c/32\pi^2$.
For the completeness, we display also the soft-momentum expansion for
functions $I_{G}(Q), \Xi(Q)$,
\ba
I_{G}(Q) &=& \frac{N_c}{8\pi^2}
\left(\ln\frac{\Lambda^2}{m^2} -1 + \widetilde\xi
+ \frac{<G^2_{\mu\nu}>}{24 N_c m^4}- \frac{Q^2}{m^2}\left(\frac16 +
\frac{<G^2_{\mu\nu}>}{40 N_c m^4}\right)\right);\nl
\Xi(Q) &=& - \frac{1}{8\pi^2} \frac{<G^2_{\mu\nu}>}{24 m^4} +
\frac{Q^2}{8\pi^2 m^2} \frac{<G^2_{\mu\nu}>}{60 m^4}.
\ea
Thus one can see that the relations (\ref{l12}) cannot be fulfilled
for any reasonable choice of $\widetilde\xi$ (which gives
$I(0) \not=0$).

Let us refer our analysis to the identities (199)-(201)
in the paper [8]  for one-fermion loop (OFL) polarization operators,
scalar, $\bar\Pi_S$, pseudoscalar, $ \bar\Pi_P$,
and longitudinal axial-vector one, $ \bar\Pi^{(0)}_A$. They are
valid in the non-gauged NJL model. The PCAC relations (199),(200) connect
the OFL pseudoscalar correlator, $\bar\Pi_P$,
 with the OFL longitudinal axial-vector one, $ \bar\Pi^{(0)}_A$.
In the GNJL model only these two relations survive.
Meantime, Eq.(201) is broken due to gluon corrections.
Indeed,
as a consequence of (\ref{l10}),(\ref{BRZ}) one has,

\ba
\bar\Pi_S &=& -\frac{< \bar q q>}{m} - (Q^2 + 4m^2) I_G(Q^2)
- Q^2 \Xi(Q^2) +O(\frac{1}{\Lambda^2}),\nl
\bar\Pi_P &=& -\frac{< \bar q q>}{m}
- Q^2 ( I_G(Q^2) + \Xi (Q^2)) + O(\frac{1}{\Lambda^2}).
\ea
The axial Ward identity (for $\xi = 0$) allows to find $\bar\Pi^{(0)}_A(Q)$,
\ba
4m^2\bar\Pi_P(Q) =&& -4m < \bar q q> + Q^4 \bar\Pi^{(0)}_A(Q);\nl
\bar\Pi^{(0)}_A(Q) =&&  - \frac{4m^2}{Q^2}
(I_G(Q^2) + \Xi (Q^2)).
\ea
Therefore the discrepance in eq.(201) of [8] is solely due to non-zero
$\Xi$,
\be
\bar\Pi_P(Q)  + Q^2 \bar\Pi^{(0)}_A(Q) - \bar\Pi_S(Q)
= - 4m^2 \Xi(Q^2) \not= 0.
\ee
In the soft-momentum
expansion, ($\xi =0$), one obtains,
\ba
- \bar\Pi^{(0)}_A(Q) &=&  \frac{N_c}{2\pi^2}
\left[\frac{1}{Q^2}\left(m^2\left(\ln\frac{\Lambda^2}{m^2} -1\right) +
\frac{<G^2_{\mu\nu}>}{24 N_c m^2}\right) - \left(\frac16
+ \frac{<G^2_{\mu\nu}>}{120 N_c m^4}\right)\right];\nl
 4m^2\bar\Pi_P(Q) &=& - 4m < \bar q q> + Q^4 \bar\Pi^{(0)}_A(Q);\nl
\bar\Pi_P(Q)  &+& Q^2 \bar\Pi^{(0)}_A(Q) - \bar\Pi_S(Q)
= \frac{<G^2_{\mu\nu}>}{48 \pi^2 m^2} + O ({Q^2\over m^2}) \not= 0.
\ea
Thus these identities prove that the results of [8] should be
revised.

\bigskip

\noindent {\bf 6. Determination of meson masses}

\bigskip

\noindent The invalidity of the $\sigma$-model relations is of course
a bad news since we cannot calculate reliably two-point correlators
in the GNJL model for finite momenta
$\sim 1 GeV$ but would like to
obtain the information about spectra from the soft-momentum expansion.
However, it is not a property of the Chiral Symmetry Breaking
to provide the linear $\sigma$-model relations. The QCD-motivated pattern
of the CSB [7] includes the radial excitations of pions and scalars with
unequal masses and coupling constants that is an extra source of
the CSB. Two interpolation schemes of mass spectrum determination can be
developed.

\bigskip

\noindent
1.\quad Following the planar limit of the QCD, eq.(\ref{planar}) one can
make the
two-resonance ansatz for scalar and pseudoscalar correlators in the
GNJL model, eq.(\ref{Wact}),
\ba
\Pi^S (Q) &=& \frac{<\bar q q>^2}{m^2}  \Gamma_{\sigma}^{-1} +
\frac{<\bar q q>}{m} =
\frac{Z^{\sigma}_0}{Q^2 + M^2_{\sigma}} + \frac{Z^{\sigma}_1}{Q^2
+ M^2_{\sigma'}} + C^{\sigma};\nl
\Pi^P (Q) &=& \frac{<\bar q q>^2}{m^2}  \Gamma_{\pi}^{-1} +
\frac{<\bar q q>}{m} =
\frac{Z^{\pi}_0}{Q^2}+
\frac{Z^{\pi}_1}{Q^2 + M^2_{\pi'}}  + C^{\pi}. \la{ans}
\ea
We remark that for the NJL-type models the constants $C_1^{S,P} = 0$.
>From the requirement of asymptotic CS restoration (\ref{CSR}) it follows
that,
\ba
C^{\sigma} &=& C^{\pi} \equiv C = \frac{<\bar q q>}{m} < 0;\\
Z^{\sigma}_0 &+& Z^{\sigma}_1 = Z^{\pi}_0 + Z^{\pi}_1;\\
Z^{\sigma}_0 M^2_{\sigma} &+& Z^{\sigma}_1 M^2_{\sigma'}
=  Z^{\pi}_1 M^2_{\pi'}.
\ea
The first two relations can be fulfilled in the conventional NJL model
which corresponds to the one-resonance ansatz, $Z_1^{\sigma,\pi} = 0$,
whereas the last one can be saturated only in a two-resonance model
including at least $\pi'$-meson.

The soft-momentum limit of the correlators is connected to the structural
constants of the chiral lagrangian [8, 13],
\ba
\frac{Z^{\sigma}_0}{M^2_{\sigma}} &+& \frac{Z^{\sigma}_1}{M^2_{\sigma'}}
 + C = 8B^2_0 (2L_8 + H_2);\nl
\frac{Z^{\pi}_1}{M^2_{\pi'}}  &+& C =  8B^2_0 (- 2L_8 + H_2).
\ea
When eliminating the unobservable constant $H_2$ one comes to the
large-$N_c$ sum rule for meson parameters based on the phenomenological
constant $L_8$ [13],
\be
\frac{Z^{\sigma}_0}{M^2_{\sigma}} + \frac{Z^{\sigma}_1}{M^2_{\sigma'}}
 - \frac{Z^{\pi}_1}{M^2_{\pi'}}  =  32 B^2_0 L_8 .
\ee
Let us interpolate now the soft-momentum expansion (\ref{sm1}) for
$\Gamma_{\sigma,\pi} $  of the GNJL model, eq.(\ref{sme}) by means of
the two-resonance ansatz (\ref{ans}),
\ba
a^{\sigma}_0 &=& C^2 \frac{M^2_{\sigma} M^2_{\sigma'}}
{Z^{\sigma}_0 M^2_{\sigma'} + Z^{\sigma}_1 M^2_{\sigma}};\quad
a^{\sigma}_1 = C^2 \frac{Z^{\sigma}_0 M^4_{\sigma'} +
Z^{\sigma}_1 M^4_{\sigma}}{Z^{\sigma}_0 M^2_{\sigma'} +
Z^{\sigma}_1 M^2_{\sigma}};\nl
a^{\pi}_1 &=& C^2 \frac{1}{Z^{\pi}_0}; \quad
a^{\pi}_2 = - C^2 \frac{Z^{\pi}_1}{\left(Z^{\pi}_0\right)^2 M^2_{\pi'}}
\ea
As a consequence, seven meson parameters $M_{\sigma, \sigma', \pi'};\quad
Z^{\sigma, \pi}_{0,1}$ are determined by entries of the GNJL models,
$\Lambda, m, <G^2_{\mu\nu}>$. The preliminary estimates show that
the gluon condensate of the GNJL model should be taken less than the
similar nonperturbative parameter in the high-energy OPE expansion [11,12].
It is not surprising since partially the nonperturbative gluon effects
are taken into account in the four-quark interaction.

\bigskip

\noindent
2.\quad In the recent literature there  appeared few conjectures [4,8,10] of
the validity of $\sigma$-model relations like
$I_{\pi} = I_{\sigma}$ (see eq.(\ref{l3}))  and $m_{\sigma}= 2m_{dyn}$
in the case of a NJL model coupled with external vector fields. In particular,
in the review [10] the influence of external electric or magnetic fields
with constant strength and direction was considered for the case of
two-point correlators.
These configurations seem to
form the same background condensate
 as gluons at the lowest order in fields and external momenta.
However it can be checked that
the restorating of the Lorentz symmetry of the vector-field
condensate (by averaging over Lorentz rotations) mixes the longitudinal
and  transversal (to the external momentum)  components.
It results in
the breaking of linear $\sigma$-model relations (\ref{l3}).

Nevertheless one can develop the soft-momentum scheme for the determination
of scalar meson masses based on the approximation,
\ba
\Gamma_{\sigma} &\simeq& q^2 \left(I_{G}(0) + \Xi(0)\right) + 4m^2 I_{G}(0);\quad
\Gamma_{\pi} \simeq q^2 \left(I_{G}(0) + \Xi(0)\right);\nl
4m^2_{\sigma} &\simeq& \frac{4m^2 I_{G}(0)}{I_{G}(0) + \Xi(0)} =
\frac{a_0^{\sigma}}{a_1^{\pi}},
\ea
that should be compared with adiabatic $\sigma$-model relations (\ref{l7}).
The Nambu relation is not valid in this scheme,
$m^2_{\sigma} > 4m^2$.

If to append the two-resonance ansatz to the adiabatic approximation one
arrives to the remarkable condition for scalar meson masses,
\be
m^2_{\sigma} = m^2_{\sigma'} \frac{Z^{\pi}_0 - Z^{\sigma}_0}{Z^{\sigma}_1},
\ee
which guaranties the validity of above approximation.
As $Z^{\pi}_0 > Z^{\sigma}_0$ one expects that in the chiral limit
the pion and $\sigma$-meson decay constants are different,
$F_{\pi} < F_{\sigma}$.

\bigskip

{\it Acknowledgements} \quad We would like to thank J. Bijnens and
M.K. Volkov for discussions as well as V.L.Yudichev for useful remarks.

\end{document}